# Quantitative Interleaved-Pulse Acquisition Extends the Range of Isotope-Ratio Measurements in ToF-SIMS

Anton V. Ievlev

*Center for Nanophase Materials Sciences, Oak Ridge National Laboratory, Oak Ridge, Tennessee 37831, United States*

## ABSTRACT

Isotope ratios underpin tracer studies of transport in oxides, battery materials, and biological systems, yet time-of-flight secondary ion mass spectrometry (ToF-SIMS) struggles to quantify ratios whose members differ in intensity by orders of magnitude: a primary-ion pulse long enough to count the minor isotope precisely can saturate the major isotope, whereas a pulse short enough to keep the major isotope linear yields too few minor-isotope counts. We show that multi-pulse-width acquisition yields quantitative isotope ratios when a dose factor measured within each crater links the phases and each isotope is taken only from phases in which it remains linear; we call this quantitative reconstruction interleaved-pulse acquisition. Pulse widths are selected from measured transfer curves and peak shapes. On thermal $SiO_2$, a tuned (6, 35) ns pair achieves the same per-layer precision as short-pulse acquisition with a 13-fold reduction in the number of analysis frames. On $^{18}$O-enriched $WO_x$ films, interleaving extends the quantifiable range at the low-fraction end by about 14-fold. Across isotope ratios spanning approximately 1:500 to 1:1, the interleaved reconstruction agrees with the corresponding linear reference measurements, showing no measurable bias introduced by interleaving, and natural-abundance ratios of oxygen, strontium and titanium are reproduced within several percent. The method requires no hardware additions and enables depth-resolved isotope-ratio measurements over a wide range within a single crater.

Time-of-flight secondary ion mass spectrometry (ToF-SIMS) is a standard tool for isotope ratios and their spatial distribution, with applications ranging from $^{18}O$ and $^{6}Li/^{7}Li$ tracer studies of ionic transport in oxides[1-5] and battery materials[6, 7] to isotope labeling in biology[8, 9] and nuclear materials.[10, 11] Ratios are attractive for quantification because isotopes of the same element share essentially the same chemistry, so many of the matrix- and yield-dependent effects that complicate quantitative SIMS cancel. Quantifying ratios, however, often requires measuring signals that differ by orders of magnitude. The natural $^{18}O/^{16}O$ ratio is about 1:500, and $^{18}O$ tracer studies follow fractions from this background up to tens of percent within one depth profile. With single-ion counting, the measurable range is capped at the top by detector saturation of the major signal and at the bottom by counting statistics of the minor one. A primary-ion pulse long enough to count the minor isotope well saturates the major; a pulse short enough to keep the major linear leaves the minor starved of counts.

Several approaches have been used to resolve this conflict, each with limitations. Computationally, the counting losses of the major signal can be corrected with Poisson statistics.[12] This method extends the usable range, but the correction diverges as the detector approaches full occupancy, so a strongly saturated signal cannot be recovered and the validity range must be checked for each measurement.[13, 14] Experimentally, a single compromise pulse width sacrifices both ends of the usable range and can leave only a narrow operating window between insufficient minor-isotope statistics and saturation of the major isotope. Hardware solutions either reduce the intensity delivered per pulse, as in burst mode and collimated burst alignment,[1, 15] at the cost of minor-isotope counts, or selectively attenuate the dominant signal,[16] which requires an additional analyzer option (extended dynamic range, EDR) not available on many instruments. Recent instrument software (SurfaceLab, IONTOF) includes a high-dynamic-range (HDR) mode that records two pulse durations (or beam currents) in one run and merges them into saturation-free images and profiles[17] by analogy with HDR photography.[18] In our tests, the merged data behaved as a superposition controlled by global coefficients rather than by a dose factor measured in each analysis. When one phase was saturated, the merged intensity was not dose-consistent and did not preserve quantitative isotope ratios under the tested conditions (Supporting Information, S7).

Sequential measurements under different conditions provide another possibility, but they multiply the measurement time and, because ToF-SIMS is destructive, do not interrogate the same material. This is particularly important for small or unique specimens such as individual exfoliated flakes, devices, or grains, for which nominally equivalent specimens may differ in thickness, size, or composition, so their data cannot simply be combined. Longer integration at a single short pulse width (short-pulse acquisition) also carries a direct cost in depth profiling: precision improves only as the square root of the collected counts, so a condition that yields ten times fewer minor-isotope counts per frame requires about ten times more acquisition for the same counting precision. This consumes proportionally more specimen per data point or requires thicker sampled layers, reducing depth resolution. The relevant figure of merit is therefore precision per sputtered layer: how precisely a ratio can be determined from a given amount of specimen.

Acquisition at multiple pulse widths within a single analysis cycle is available on current instruments and has previously been used for Cs-cluster detection.[19] What has been missing is a reconstruction that makes the combined data quantitative. Here we introduce interleaved-pulse acquisition as such a method for isotope ratios that span orders of magnitude within a single depth profile. Each analysis cycle contains two or more primary-ion pulse widths: a short phase keeps the major isotope within the linear counting regime, a long phase provides improved counting statistics for the minor isotope, and a dose factor measured within each crater links the phases, with each isotope taken only from the phase in which it is linear. The method requires no hardware additions. We choose pulse widths from the measured transfer curve and peak shapes and evaluate the method on thermal $SiO_2$, a $SrTiO_3$ single crystal, and $^{18}O$-enriched $WO_x$ films. Relative to short-pulse acquisition, interleaving achieves the same per-layer precision with a 13-fold reduction in the number of frames per layer and extends the quantifiable range about 14-fold at the lower end. Across isotope ratios from 1:500 to about 1:1, the interleaved reconstruction agrees with the linear short-pulse measurements, adding no measurable bias, and natural-abundance ratios of oxygen, strontium and titanium are reproduced within several percent.

# EXPERIMENTAL SECTION

## *Instrument and acquisition*

Measurements were performed using a TOF.SIMS 5-NCS instrument (IONTOF GmbH, SurfaceLab 7.5) without the extended-dynamic-range option, operated in spectrometry (bunched) mode and negative polarity unless stated otherwise. A pulsed $Bi_3^+$ beam was used as the analysis source (30 keV, 30 nA DC current, 5 µm spot size) and rastered over a 100 × 100 µm² area at 128 × 128 resolution with one shot per pixel. Pulse widths between 1 and 60 ns were used. Depth profiles were acquired in the instrument's non-interlaced depth-profiling mode: each analysis cycle, containing all interleaved phases, was followed by $Cs^+$ sputtering (1 keV, 80 nA, 300 × 300 µm²) for 3.06 s, followed by a 2 s pause, so that one cycle corresponds to one depth layer. In interleaved-pulse acquisition the phases of a cycle are acquired in sequence, short phase first, and the dose of each phase is set by the number of frames per scan. Switching between phases adds 122 ± 7 ms per phase and scan. A low-energy electron flood gun was used in all measurements; for $SrTiO_3$ in negative polarity the extractor bias was additionally corrected for sample charging. Positive-polarity $SrTiO_3$ measurements used the same conditions without the extractor-bias correction.

## *Samples*

Thermal $SiO_2$ (100 nm dry oxide on Si) was taken from sealed wafer stock. A $SrTiO_3$(100) single crystal (Alfa Aesar/Thermo Fisher Scientific, product 38507, lot Z15C035; 10 × 10 × 1 mm³, one side polished, Verneuil-grown, undoped) served as a second natural-abundance sample. Four $WO_x$ films were grown on thermally oxidized Si by reactive DC magnetron sputtering in $Ar/O_2$ (growth details in the Supporting Information). A buried isotopic marker was produced by replacing part of the $^{16}O_2$ flow with $^{18}O_2$ at constant total oxygen flow, giving nominal marker fractions of 50, 25, 16.7 and 12.5% assuming isotopically pure $^{18}O_2$. The films are nominally 45 nm thick with the ~15 nm $^{18}$O-enriched layer in their center and were capped with ~30 nm $SiN_x$.

### *Data treatment*

Exported counts carry the vendor Poisson dead-time correction for a single-stop detector, applied per pixel, which we verified with the raw counts (*Figure 1c*); no further rate correction was applied. Because the peak position shifts with pulse width, integration windows were placed relative to each phase's own peak and matched between the two isotopes; the weak $^{18}$O peak was corrected for the $^{17}$OH contribution and the local continuum. The dose factor *k* was computed for each crater from its steady-state layers, after the first five layers of Cs equilibration (Figure 3a). Per-layer precision is the layer-to-layer scatter of the ratio after removing a linear trend. Natural isotope abundances are IUPAC values;[20] the $^{18}O/^{16}O$ reference ratio is that of VSMOW, 0.0020052.[21] The source data underlying the figures and tables, and the exported depth profiles and spectra from which they were computed, are available in the accompanying Zenodo repository.

## RESULTS

### *Interleaved-pulse acquisition*

In interleaved-pulse acquisition, each analysis cycle consists of two or more phases, each acquired with its own predefined set of acquisition parameters, followed by a sputter step (*Figure 1*a); one cycle yields one depth layer. Several parameters can differ between phases: the number of frames per scan, the number of shots per pixel, the raster resolution and the primary-ion pulse width. The first three change only the number of primary-ion pulses delivered during a phase: the collected counts scale with the acquisition time, while the counts per pulse stay the same. In contrast, the pulse width changes the dose delivered by each pulse, and hence both the counting statistics and the degree of saturation, at no cost in time, since even a 60 ns pulse is a negligible part of the 75 µs pulse period. Pulse width therefore provides the primary control over the per-pulse dose and saturation state, whereas frames per scan determine the total dose allocated to each phase.

In the configuration used below, a short phase (6 ns) is acquired with three frames summed into one scan and a long phase (35 ns) with one frame. The short phase measures the major isotope, $^{16}$O, in the linear counting regime; the long phase provides the minor isotope, $^{18}$O, with far more counts. The two phases are linked by a crater-specific dose factor $k$, determined from the minor isotope over the steady-state portion of the profile:

$$k = \sum {}^{18}O_{long} / \sum {}^{18}O_{short} \quad (1)$$

where the sums run over the steady-state layers of the crater and $^{18}O$ is linear in both phases. The $^{18}O$ fraction of each layer then follows from its per-layer counts as:

$$f = \frac{{}^{18}O_{long}}{\left(k \cdot {}^{16}O_{short} + {}^{18}O_{long}\right)} \quad (2)$$

Both phases probe the same layer within the same cycle. Changes in secondary-ion yield, primary-ion current or charging along the profile therefore affect both phases alike and cancel in $f$. What differs between the phases is their dose ratio, including any dependence of the transmission on pulse width, and this is captured by $k$, measured in every crater. The reconstruction thus needs no stored calibration and holds as long as the dose ratio stays constant within the crater. The approach can be extended to more than two phases, provided that each phase shares with the next an isotope of the same element that is linear in both (the dose factor depends on mass); the extension to any number of phases is given in the Supporting Information (S6).

### *Choosing the pulse widths*

With the acquisition scheme defined, the practical question is how to choose the pulse width for each phase. Two constraints set the pulse widths: the short phase must keep the major isotope in the linear counting regime, and the long phase must keep the minor-isotope peak separated from its neighbors. Because the response of both the count rate and the peak shape to the pulse width is strongly nonlinear, we first measured the transfer curve on thermal $SiO_2$ for 21 pulse widths from 1 to 60 ns (*Figure 1b*). The $^{18}O$ rate rises monotonically from $4.6 \times 10^{-5}$ to 0.21 counts per pulse, and stays far from saturation over the whole range. $^{16}O$ follows it up to 8 ns and then saturates near 6 counts per pulse.

The linear range of $^{16}O$ thus appears to extend to several counts per pulse, far beyond the ≈ 0.8 counts per pulse threshold often used in practice. This is because the vendor software corrects the counts for detector dead time. To test the correction near saturation, we plot the $^{16}O/^{18}O$ ratio against the corrected $^{16}O$ count rate (*Figure 1c*). Since $^{18}O$ remains linear, any error in the $^{16}O$ correction would appear as a change of the ratio. The ratio stays constant from 0.02 to 3.6 counts per pulse (1–8 ns), which verifies the correction over this range.[14] The raw counts show what the correction does: uncorrected, the ratio would already read 29% low at 6 ns and 75% low at 8 ns, as predicted by the Poisson dead-time model for a single-stop detector, $\left(1-e^{-x}\right)/x$, where $x$ is the true (corrected) counts per pulse.[12] We place the short phase at 6 ns, where the detector registers a count on about half of the pulses and the correction factor is about 1.4; beyond about 7 ns the correction approaches its singularity and becomes fragile.

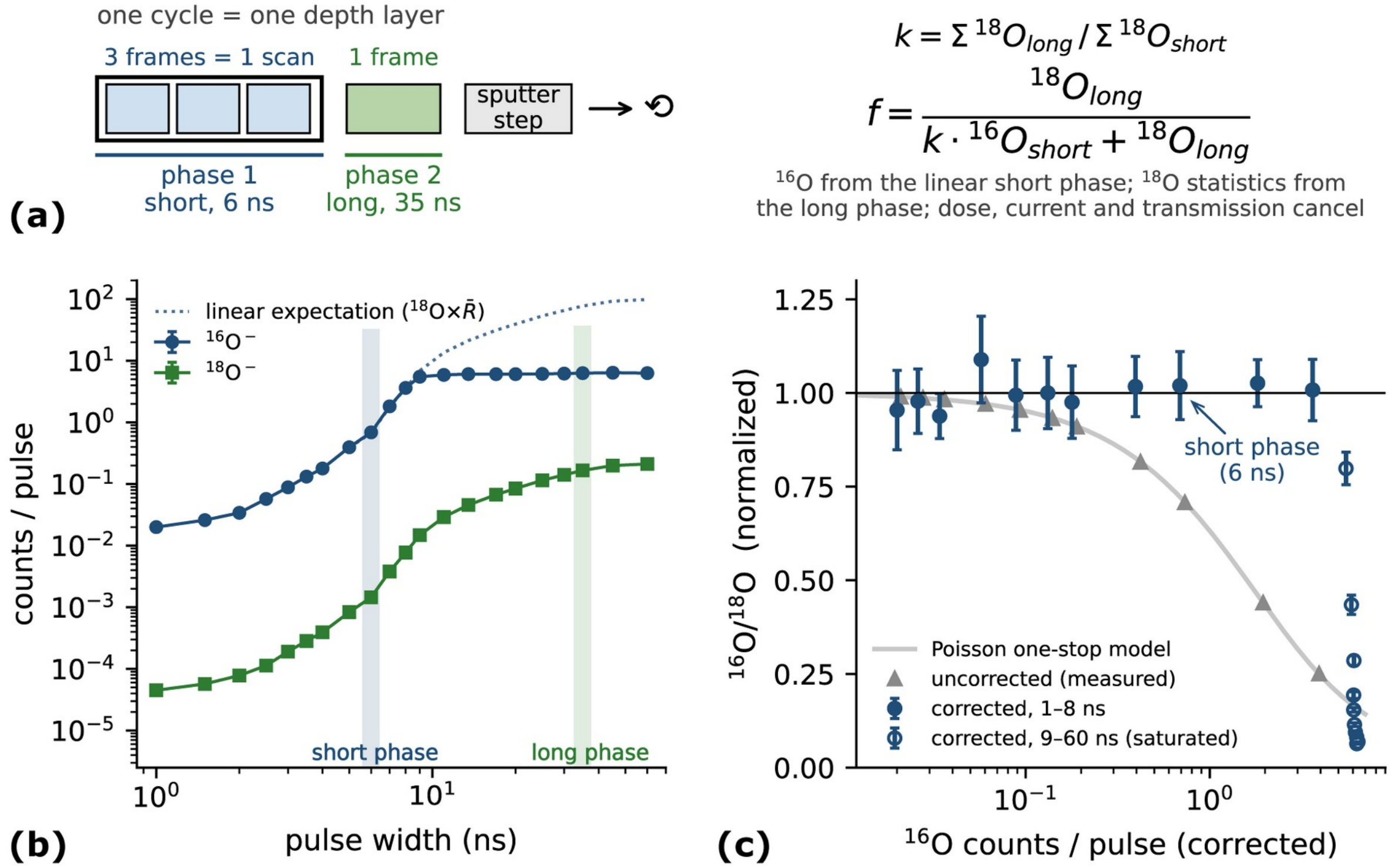

***Figure 1.*** *(a) Acquisition cycle in interleaved-pulse acquisition: two analysis phases (6 ns, three frames per scan; 35 ns, one frame) followed by the sputter step. (b) Transfer curve on thermal* $SiO_2$*: counts per pulse of* $^{16}O^-$ *and* $^{18}O^-$ *vs pulse width (mean ± SD of four craters). Dotted line: linear expectation for* $^{16}O^-$ *(signal of* $^{18}O^-$ *scaled by the mean* $^{16}O^-/^{18}O^-$ *ratio of the linear range). Shaded bands: selected short and long phases. (c)* $^{16}O^-/^{18}O^-$ *ratio, normalized to its mean over 1–8 ns, vs the dead-time-corrected* $^{16}O^-$ *count rate. Triangles: the same ratio without dead-time correction, from the raw counts of one crater. Grey line: Poisson dead-time model for a single-stop detector.*

The long phase is different: it is limited not by counting, since $^{18}O$ stays below 0.25 counts per pulse at every width, but by what the pulse width does to the peak shape and width (*Figure 2*). Short pulses give broader peaks with a long high-mass tail (FWHM 5.7 mDa at 1 ns); at the 6 ns short phase the $^{16}O^-$ peak is 3.7 mDa wide at half maximum and 16 mDa at 1% of maximum (*Figure 2a*). The $^{18}O^-$ peak keeps a compact shape up to 35 ns; beyond it, ions outside the buncher acceptance form a low-mass pedestal that reaches about 5% of the maximum at 60 ns (*Figure 2b*). The shoulder at +7 to +11 mDa, about 1% of the $^{18}O^-$ peak, is the unresolved $^{17}OH^-$ and $H_2O^-$ signal.

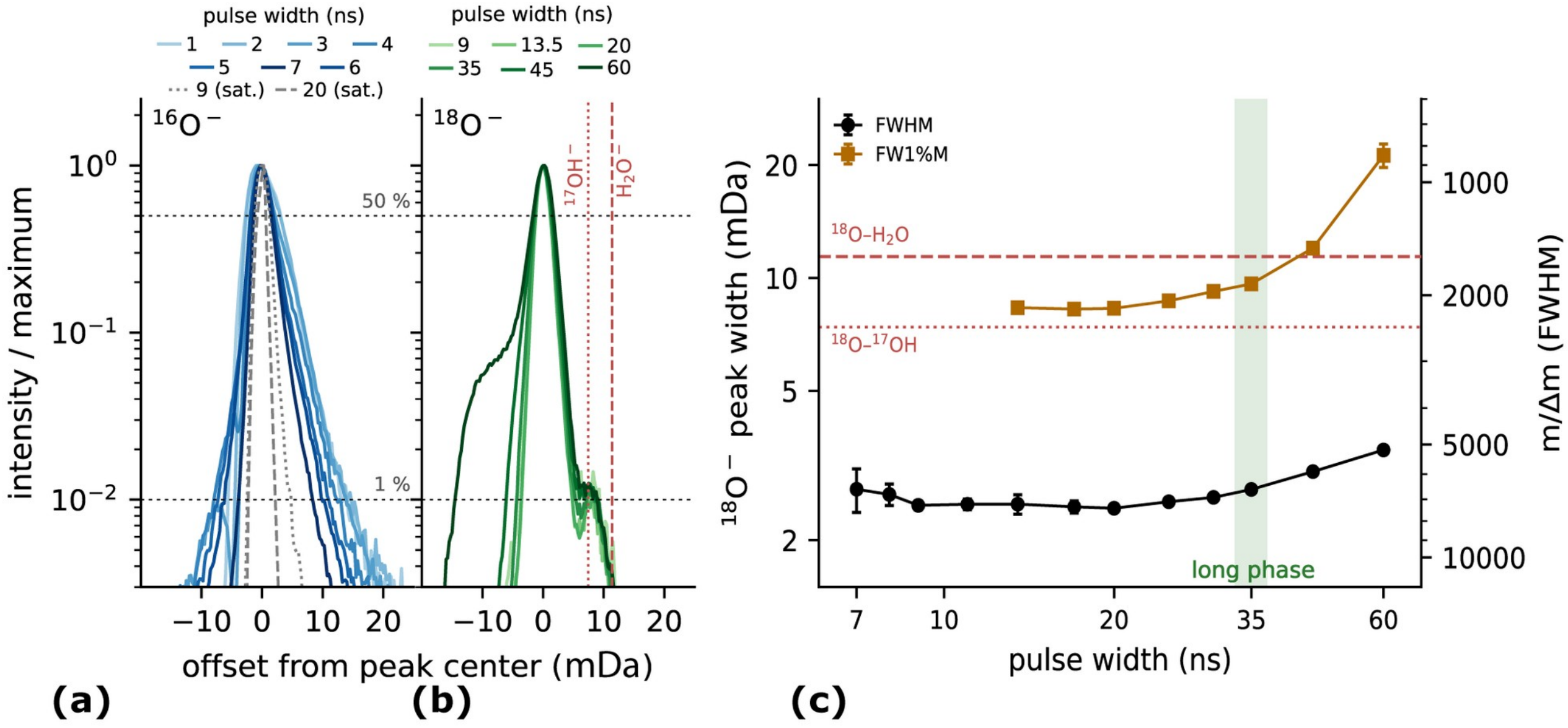

***Figure 2.*** *Effect of the pulse width on the peak shape in thermal $SiO_2$. (a, b) Peak shapes of $^{16}O^-$ at 1–7 ns and $^{18}O^-$ at 9–60 ns (four craters, each centered on its own peak before summing; log scale, normalized to the maximum). Grey curves in (a): saturated $^{16}O^-$, which appears artificially narrow because the dead-time correction distorts the peak core. Dotted and dashed lines in (b): positions of the $^{17}OH^-$ and $H_2O^-$ neighbors. (c) Width of the $^{18}O^-$ peak against pulse width (mean ± SD of four craters): FWHM (right axis: equivalent m/Δm) and full width at 1% of maximum (FW1%M), shown where the peak contains sufficient counts for the level measured (≥50 counts at the apex for FWHM, ≥1000 for FW1%M). Horizontal lines: spacing to the $H_2O^-$ and $^{17}OH^-$ neighbors; a peak is separated from a neighbor of the same line shape at a given level when its full width at that level lies below the spacing. Shaded band: selected long phase.*

The width of the $^{18}O^-$ peak (*Figure 2c*) sets the pulse width of the long phase. From 8 to 35 ns the $^{18}O^-$ peak is 2.4–2.7 mDa wide at half maximum (*m*/Δ*m* 6600–7400), and from 13.5 to 35 ns its full width at 1% of maximum (FW1%M) stays at 8.3–9.6 mDa, inside the 11.4 mDa spacing to the $H_2O^-$ peak. Because neighboring peaks are expected to share the same instrumental line shape, FW1%M provides a direct criterion for their separation at the 1% level. At 45 ns the FW1%M reaches 12.0 mDa, and at 60 ns the pedestal raises it to 21 mDa. We therefore take 35 ns, the longest width whose FW1%M stays inside the $H_2O^-$ spacing. The 6 ns short phase is not separated from $H_2O^-$ at the 1% level, but on a dry oxide the $H_2O^-$ signal is below 1% of $^{18}O^-$, so the overlap is negligible. Finally, the short phase receives three frames per scan and the long phase one, because the $^{18}O$ counts of the short phase limit the precision of both the dose factor *k* and the crater-averaged ratio.

### *Performance*

With the pulse widths selected, we next compare the tuned (6, 35) ns pair with short-, compromise-, and long-pulse single-width acquisitions, as well as with an untuned interleaved pair, (6, 9) ns. The strategies are compared by their per-layer precision, the layer-to-layer scatter of the ratio, because it sets the noise of a depth profile and how far a diffusion tail can be followed. All strategies use four analysis frames per cycle and the same sputter step, so a gain in precision translates directly into less analysis time per layer or, at equal precision, proportionally thinner layers and less consumed material (*Table 1*). Note that this gain applies to individual layers in a depth profile rather than to the profile-integrated ratio. Because $k$ normalizes the long phase to the short phase, the ratio integrated over the entire measurement reduces to that of the short phase alone and therefore does not benefit from interleaving. The same counting-statistical argument should also apply to spatially resolved isotope ratios in imaging, where the relevant unit is a pixel rather than a depth layer.

***Table 1.*** *Precision and time per layer for acquisition strategies on thermal $SiO_2$ ($^{18}O/^{16}O$, 4 analysis frames per cycle). Precision is the measured per-layer scatter of the ratio; deviation is relative to the ratio measured without saturation (its offset from natural abundance is given in Table 2). Frames and time per layer are those required to reach 2.6%, the precision of the tuned pair; time includes 3.06 s sputtering and a 2 s pause per layer. The full table is given in the Supporting Information.*

| Strategy | Precision per layer (4 frames) | Deviation from unsaturated ratio | Frames per layer for 2.6% precision | Time per layer for 2.6% precision |
| --- | --- | --- | --- | --- |
| short pulse only, 6 ns | 9.3% | 0 | 51 | 67 s |
| compromise, 8 ns | 4.5%[a] | −1% | 12[a] | 19.5 s |
| single width, 9 ns | 2.5% | +21 to +25% | 3.6 | 9.5 s |
| untuned pair, (6, 9) ns | 3.2% | 0 | 6 | 12.9 s |
| **tuned pair, (6, 35) ns** | **2.6%** | **0** | **4** | **10.2 s** |
| long pulse only, 35 ns | 1.6–2.1%[b] | +1100% | 1.5–2.5[b] | 7.0–8.3 s |

[a] Predicted from the transfer curve. [b] Measured with one frame and scaled to four frames, assuming that the additional noise from the saturated-$^{16}O$ correction averages over frames; counting statistics alone predict 1.0% at one frame.

The comparison (*Table 1*) reveals a substantial reduction in the time required to reach a given precision. Relative to short-pulse acquisition, the tuned pair shows a 13-fold reduction in the number of frames required to reach the same 2.6% precision. At a fixed sputter increment, this gain reduces the analysis time required per layer; alternatively, it can be traded for finer depth sampling by reducing the sputter dose per cycle while maintaining the same per-layer precision. This corresponds to a 12.2-fold reduction in analysis time and a 6.6-fold reduction in total time per layer, including the common sputter step and pause.

A single compromise width cannot deliver this. At 8 ns the ratio is still within 1% of the unsaturated value, but the detector is then occupied on about 98% of pulses; at 9 ns the ratio is already 21–25% high, and at 35 ns it is 12 times too high. The saturated options buy almost no time in return: the 9 ns and 35 ns single widths are only 7% and 20–30% faster than the tuned pair. Saturation also costs precision, since the correction of the saturated $^{16}O$ adds 2.5–3.7% per layer of its own noise.[22] For comparison, the untuned pair (6, 9) ns, chosen without measuring the transfer curve, reaches 3.2% and needs 1.3 times the time of the tuned pair; tuning the widths is thus a refinement, and most of the gain comes from interleaving itself.

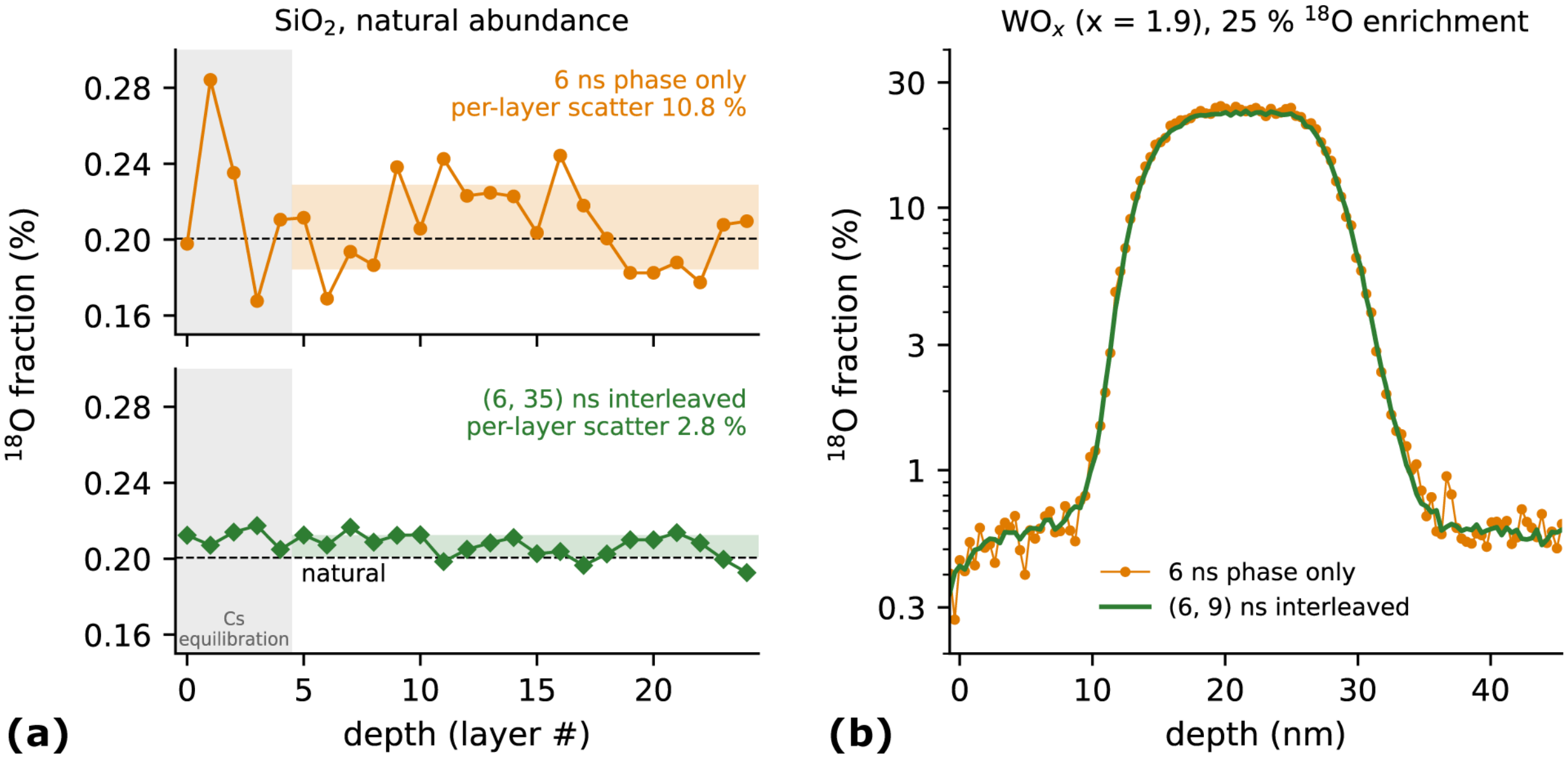

***Figure 3.*** *Per-layer noise in representative depth profiles. (a) Thermal $SiO_2$, one crater: $^{18}O$ fraction per layer from the 6 ns phase alone (top; three frames per layer; Table 1 gives the four-frame equivalent) and from the (6, 35) ns interleaved reconstruction (bottom), plotted on identical scales. Bands: mean ± 1σ of the per-layer scatter (layers 5–24, linear trend removed). Grey region: Cs equilibration, excluded from the statistics. Absolute levels are referenced to the crater-averaged ratio obtained using matched integration windows. (b) $WO_x$ film with 25% nominal $^{18}O$ enrichment: 6 ns phase alone compared with the (6, 9) ns interleaved reconstruction at each layer. The depth scale is nominal (45 nm film over 119 layers). The flank level of ~0.6% is a property of the as-grown film.*

*Figure 3a* shows what these numbers mean in a real profile, measured in thermal $SiO_2$. Both traces come from the same crater: the 6 ns phase alone scatters by 10.8% per layer (9.3% scaled to four frames), the interleaved reconstruction by 2.8%. During the first layers, while the surface reaches Cs equilibrium, the $^{16}O$ yield rises about 13-fold; the reconstruction stays within ±5% of its steady-state value throughout, whereas the ratio of the short phase alone swings from −19% to +38%. Single phases follow counting statistics within 1.0–1.1×, including repeated identical phases within a cycle. The tuned pair does not quite: its per-layer scatter, 2.6% (Table 1), is slightly larger than the 2.0% expected from its $^{18}O$ counting statistics alone; the difference corresponds to an additional contribution of approximately 1.6% in quadrature. This contribution does not depend on the counts and sets a practical floor: once it is reached, additional long-phase counts provide little improvement, and extra frames are better allocated to the short phase, where they improve both the dose factor and the crater-averaged ratio.

## DISCUSSION

The Results establish what interleaving gains in precision and time; here we discuss the implications for accuracy, quantifiable range, calibration, and the limits of applicability.

### *Accuracy*

To assess absolute accuracy, as distinct from the saturation bias evaluated in Table 1, we measured three sample systems of known or nominally known isotopic composition (*Table 2*). Thermal $SiO_2$ and a $SrTiO_3$ single crystal, both at natural abundance, test absolute accuracy: $SiO_2$ for $^{18}O/^{16}O$ in negative polarity, $SrTiO_3$ for $^{18}O/^{16}O$ in negative polarity and for the Sr and Ti isotope ratios in positive polarity. The $WO_x$ films with $^{18}O$-enriched markers of 12.5–50% nominal fraction test linearity at high minor-isotope fractions, where the long phase approaches saturation.

On thermal $SiO_2$ the $^{18}O/^{16}O$ ratio reads 0.9 ± 2.1% above the VSMOW value (seven craters), and on $SrTiO_3$, which required additional charge compensation, 3.4 ± 2.3% above it. In positive polarity, $^{86}Sr/^{88}Sr$ and $^{87}Sr/^{88}Sr$ agree with natural abundance within +3.4 and +2.7%, and the Ti ratios within −3.5 to +6.8%. The largest deviations occur where isobaric interferences are expected, $^{48}TiH^+$ on $^{49}Ti$ (TiH/Ti = 0.62%) and possible $^{46}Ca^+$, $^{50}Cr^+$ and $^{50}V^+$ on $^{46}Ti$ and $^{50}Ti$, and for the lowest-abundance isotope, $^{84}Sr$ (+6.3%); $^{87}Rb^+$ on $^{87}Sr^+$ is negligible, as the $^{85}Rb^+$ monitor shows. Natural-abundance ratios are thus reproduced within several percent. Even so, absolute ratios in SIMS require a matrix-matched reference: the choice of integration window alone moves the oxygen values by ±2%, whereas normalized quantities such as tracer profiles are insensitive to such offsets. The Sr ratios are also a real-data example of a chain of phases: $^{88}Sr$ is taken from the 9 ns phase and the weak Sr isotopes from the 20 ns phase, with $^{86}Sr$, which is linear in both, linking the phases. As an internal check, $^{88}Sr$ at 20 ns also remains below 3 corrected counts per pulse and can serve as an alternative bridge; its dose factor differs by about 3%, a discrepancy reported but not resolved in the Supporting Information (S5.2 and S6.4).

***Table 2.*** *Expected and measured isotope ratios (minor/major) from interleaved-pulse acquisition. Expected natural-abundance ratios are IUPAC values; $^{18}O/^{16}O$ is the VSMOW ratio. Deviation is relative to the expected value and measures absolute accuracy; it is distinct from the deviation in Table 1, which is relative to the unsaturated ratio of the same measurement and measures saturation bias only. For the $WO_x$ markers, the note gives that latter quantity.*

| Sample | | Isotope pair | Expected | Measured | Deviation | Note |
|---|---|---|---|---|---|---|
| **$SiO_2$** | thermal oxide, 100 nm | $^{18}O/^{16}O$ | 0.002005 | 0.002023 ± 0.000042 | +0.9% | 7 craters, 2 days |
| **$SrTiO_3$** | negative polarity, extractor-bias compensation | $^{18}O/^{16}O$ | 0.002005 | 0.002073 ± 0.000046 | +3.4% | 4 craters |
| | positive polarity | $^{84}Sr/^{88}Sr$ | 0.00678 | 0.00721 ± 0.00005 | +6.3% | lowest-abundance isotope |
| | | $^{86}Sr/^{88}Sr$ | 0.1194 | 0.1235 ± 0.0007 | +3.4% | |
| | | $^{87}Sr/^{88}Sr$ | 0.0848 | 0.0871 ± 0.0004 | +2.7% | $^{85}Rb^+$ monitor ≈ 0 |
| | | $^{46}Ti/^{48}Ti$ | 0.1119 | 0.1162 ± 0.0003 | +3.8% | $^{46}Ca^+$ isobar possible |
| | | $^{47}Ti/^{48}Ti$ | 0.1009 | 0.1034 ± 0.0003 | +2.5% | |
| | | $^{49}Ti/^{48}Ti$ | 0.0734 | 0.0784 ± 0.0009 | +6.8% | $^{48}TiH^+$ interference (TiH/Ti = 0.62%) |
| | | $^{50}Ti/^{48}Ti$ | 0.0703 | 0.0678 ± 0.0006 | −3.5% | $^{50}Cr^+/^{50}V^+$ isobars possible |
| **$WO_x$** | 12.5% enrichment | $^{18}O/^{16}O$ | 0.143* | 0.1287 ± 0.0044 | —* | vs short phase |

| Sample | | Isotope pair | Expected | Measured | Deviation | Note |
|---|---|---|---|---|---|---|
| | | | | | | −1.6% |
| | 16.7% enrichment | $^{18}O/^{16}O$ | 0.200* | 0.1944 ± 0.0029 | —* | vs short phase +0.6% |
| | 25% enrichment | $^{18}O/^{16}O$ | 0.333* | 0.2993 ± 0.0049 | —* | vs short phase −1.9% |
| | 50% enrichment | $^{18}O/^{16}O$ | 1.000* | 0.8306 ± 0.0061 | —* | vs short phase +0.2% |

$^{18}O/^{16}O$: matched ±4 mDa apex windows on both isotopes, local baselines, $^{17}OH^{-}$ contribution subtracted; the uncertainty is the larger of the counting and crater standard errors; the window choice (±2.5 to ±6 mDa) moves the values by ±2%. Sr: $^{86}Sr$ bridges the 9 and 20 ns phases; $^{88}Sr$ is taken from 9 ns and the weak Sr isotopes from 20 ns, and $^{88}Sr$ at 20 ns provides an alternative bridge used as an internal check (Supporting Information, S5.2). Ti ratios are taken directly from the 20 ns phase, where all Ti isotopes are within the verified linear range. Mean ± SD of three craters. *Nominal $^{18}O$ fraction of the growth gas; the actual enrichment of the films is not known, so no deviation is given. The note compares the interleaved result with the linear short phase of the same craters.

The $WO_x$ markers test linearity rather than accuracy, since their expected values are only the nominal $^{18}O$ fractions of the growth gas and no deviation is given. The ratio of the interleaved reconstruction to the linear short phase of the same craters is 0.993 ± 0.013 over 12.5 – 50% nominal enrichment, with no trend, although the long-phase $^{18}O$ inside the markers reaches 0.4–1.6 counts per pulse. Together, these samples cover a wide range of isotope ratios from 1:500 to about 1:1. Within the validated linear ranges, the different phase sets used here primarily affect precision rather than the reconstructed ratio.

### *Quantifiable range*

Accuracy concerns the mean of a ratio; the second implication of interleaving is how wide a range of fractions can be quantified within one profile. This is the practical question in tracer work, where a diffusion profile falls from tens of percent to the natural background within a single crater. We define the quantifiable range, or dynamic range, of an isotope-ratio profile as the span of minor-isotope fractions that can be quantified at a given per-layer precision. At the low-fraction end, quantification is limited by the counting statistics of the minor isotope while the major isotope must remain linear in the short phase. At the high-fraction end, the minor-isotope signal must remain linear in the longer phase.

Interleaving extends the quantifiable range about 14-fold at the low end. At a target per-layer precision of 10%, short-pulse acquisition quantifies $^{18}O$ fractions down to about 1%, whereas the interleaved acquisition reaches about 0.07%. This inferred lower limit lies below the natural $^{18}O$ abundance of 0.2%, implying that natural abundance could be quantified layer by layer in the same profile as a ~45% marker. The corresponding quantifiable range therefore increases from about 1.6 to 2.8 decades. This is the gain of Table 1 seen from the other side: at equal time the range extends 14-fold, and at equal range the time drops 14-fold.

These limits follow from the $WO_x$ profile of *Figure 3b*, which spans two decades of $^{18}O$ fraction in one crater. At the low end, in the flanks at about 0.6% $^{18}O$, interleaving lowers the per-layer noise from 13.3 to 3.5%; at the high end, on the 22.5% plateau (25% nominal), the short phase already has ample counts and the two agree within their noise (1.9 vs 1.6%). Because per-layer noise scales approximately as the inverse square root of the fraction, a 3.8-fold reduction in noise lowers the fraction quantifiable at a given precision by approximately $3.8^2 \approx 14$-fold; extrapolating from the flank gives the limits above.

The factor of 14 is not a limit of the method but of the (6, 9) ns configuration used for these films; the range scales with the dose ratio between the phases. It can be extended by a longer pulse (35 ns gives 11 times the $^{18}O$ rate per frame of 9 ns), by more frames in the long phase, or by additional phases. A long phase tuned for dilute tails would saturate in a highly enriched marker; an intermediate phase then keeps the enriched region linear while the longest phase serves the tails, all phases forming the connected chain described above, with limits set by the peak width of the longest phase and the linearity of the shortest.

### Calibration and limits

Unlike an attenuation factor calibrated once and stored, the relative dose between phases is measured independently in every crater. Because both phases probe the same layer, common changes in secondary-ion yield, primary-ion current, or charging cancel in the reconstruction, as demonstrated by the 13-fold yield ramp during Cs equilibration (*Figure 3*a), while the crater-specific dose factor accounts for the relative dose and transmission between phases. On thermal $SiO_2$ acquired with the same 6 and 35 ns widths used in the main experiment, vendor HDR merging overestimated $^{18}O/^{16}O$ by 2.6-fold, whereas dose-calibrated reconstruction of the same component datasets agreed with the linear short-pulse crater average while reducing the per-layer scatter by about sevenfold (Supporting Information, S7). The transfer curve (*Figure 1b*) replaces assumed limits with measured ones: the usual ~0.8 counts-per-pulse threshold is not the linearity limit on this instrument, and the 8 ns compromise lies close to the onset of strong bias (*Table 1*). Together, these measurements make the approach largely self-calibrating: the relative dose is determined within each crater, while the usable pulse-width range is established experimentally for the relevant matrix, instrument, and acquisition mode.

This self-calibrating character also defines what must be verified for each application. Before a measurement, the transfer curve depends on the matrix, instrument, and acquisition mode, so pulse widths should be revalidated when these conditions change. During a profile, the dose ratio between phases can drift slowly (up to 4% over 20 layers in our data), producing a systematic bias in the reconstructed $f$ if a single $k$ is applied across the entire profile. The bridge-isotope ratio should therefore be inspected for trends over the region of interest, preferably as a running or region-averaged value when individual layers are too noisy, and k determined from layers spanning the region of interest. And because the long phase is chosen by peak width rather than by full resolution, near-doublet interferences such as $H_2O^-$ on $^{18}O^-$ are bounded rather than resolved, so the bound has to be confirmed for each matrix.

# CONCLUSIONS

Quantifying isotope ratios whose two members differ in intensity by orders of magnitude is a long-standing limitation of ToF-SIMS because no single primary-ion pulse width is optimal for both signals. Interleaved-pulse acquisition addresses this conflict by measuring the major isotope with a short pulse and the minor isotope with a long pulse within each analysis cycle, with a crater-specific dose factor linking the two and each isotope taken only from the phase in which it is linear; this, rather than the multi-width acquisition itself, is what makes the result quantitative. With pulse widths selected from measured transfer curves and peak shapes, the method achieves the per-layer precision of short-pulse acquisition with a 13-fold reduction in the number of analysis frames, introduces no measurable bias or nonlinearity relative to the linear reference over ratios from approximately 1:500 to 1:1, reproduces natural-abundance ratios within several percent, and extends the quantifiable range by about 14-fold at the low-fraction end; further gains are possible through longer phases or chains of phases.

Demonstrated here for $^{18}O/^{16}O$ in negative polarity and Sr and Ti isotope ratios in positive polarity, the approach should extend to other large-ratio measurements, including $MCs^+/Cs^+$ cluster ratios and isotopes of other elements. The same counting-statistical advantage should also extend to lateral isotope mapping, where the relevant unit is a pixel rather than a depth layer. Because acquisition parameters are programmable between phases, interleaved-pulse acquisition may also facilitate autonomous experiments in which the phase set is selected or adjusted during measurement.[19]

# ASSOCIATED CONTENT

**Supporting Information**

The Supporting Information is available. Experimental details; full performance comparison; noise-floor analysis; peak windows and interference bounds; per-crater accuracy data; extension to more than two phases; comparison with the vendor HDR mode (PDF).

## Data Availability

The source data underlying the figures and tables, and the exported depth profiles and spectra from which they were computed, are available at https://doi.org/10.5281/zenodo.23037549. Raw IONTOF project files and analysis scripts are available from the author on request.

# AUTHOR INFORMATION

## Corresponding Author

Anton V. Ievlev − Center for Nanophase Materials Sciences, Oak Ridge National Laboratory, Oak Ridge, Tennessee 37831, United States; orcid.org/0000-0003-3645-0508; Email: ievlevav@ornl.gov

### Notes

The author declares no competing financial interest.

## ACKNOWLEDGMENTS

This work was supported by the U.S. Department of Energy, Office of Science, Office of Basic Energy Sciences, Division of Materials Sciences and Engineering. ToF-SIMS measurements were conducted at the Center for Nanophase Materials Sciences (CNMS), a U.S. Department of Energy Office of Science User Facility at Oak Ridge National Laboratory. The author thanks Yiyang Li and Heather Hare (University of Michigan) for providing the $^{18}O$-enriched $WO_x$ samples.

## REFERENCES

(1) De Souza, R. A.; Zehnpfenning, J.; Martin, M.; Maier, J. Determining oxygen isotope profiles in oxides with Time-of-Flight SIMS. *Solid State Ionics* **2005**, *176* (15-16), 1465-1471. DOI: 10.1016/j.ssi.2005.03.012.
(2) Kilner, J. A.; Skinner, S. J.; Brongersma, H. H. The isotope exchange depth profiling (IEDP) technique using SIMS and LEIS. *Journal of Solid State Electrochemistry* **2011**, *15* (5), 861-876. DOI: 10.1007/s10008-010-1289-0.
(3) Holzlechner, G.; Kubicek, M.; Hutter, H.; Fleig, J. A novel ToF-SIMS operation mode for improved accuracy and lateral resolution of oxygen isotope measurements on oxides. *Journal of Analytical Atomic Spectrometry* **2013**, *28* (7), 1080-1089. DOI: 10.1039/c3ja50059d.
(4) Li, J.; Appachar, A.; Peczonczyk, S. L.; Harrison, E. T.; Ievlev, A. V.; Hood, R.; Shin, D.; Yoo, S.; Roest, B.; Sun, K.; et al. Thermodynamic origin of nonvolatility in resistive memory. *Matter* **2024**, *7* (11), 3970-3993. DOI: https://doi.org/10.1016/j.matt.2024.07.018.
(5) Shin, D.; Ievlev, A. V.; Beckmann, K.; Li, J.; Ren, P.; Cady, N.; Li, Y. Oxygen tracer diffusion in amorphous hafnia films for resistive memory. *Materials Horizons* **2024**, *11* (10), 2372-2381. DOI: 10.1039/d3mh02113k (acccessed 8/4/2026).
(6) Meyer, T.; Gutel, T.; Manzanarez, H.; Bardet, M.; De Vito, E. Lithium Self-Diffusion in a Polymer Electrolyte for Solid-State Batteries: ToF-SIMS/ssNMR Correlative Characterization and Modeling Based on Lithium Isotopic Labeling. *Acs Applied Materials & Interfaces* **2023**, *15* (37), 44268-44279. DOI: 10.1021/acsami.3c08829.
(7) Bofanova, M.; Delfino, P. M.; Audinot, J.-N.; Porcher, W.; Märker, K.; De Vito, E.; Dupré, N. Exploring Li pathways in Si-C/Gr electrodes and their SEI in Li batteries using Li isotope tracing by MAS NMR and FIB-SIMS. *Nano Energy* **2025**, *144*, 111324. DOI: https://doi.org/10.1016/j.nanoen.2025.111324.
(8) Tyler, B. J.; Takeno, M. M.; Hauch, K. D. Identification and Imaging of 15N labeled cells with ToF-SIMS. *Surface and Interface Analysis* **2011**, *43* (1-2), 336-339. DOI: 10.1002/sia.3679.
(9) Cliff, J. B.; Gaspar, D. J.; Bottomley, P. J.; Myrold, D. D. Exploration of inorganic C and N assimilation by soil microbes with time-of-flight secondary ion mass spectrometry. *Applied and Environmental Microbiology* **2002**, *68* (8), 4067-4073. DOI: 10.1128/aem.68.8.4067-4073.2002.
(10) Bosco, H.; Hamann, L.; Kneip, N.; Raiwa, M.; Weiss, M.; Wendt, K.; Walther, C. New horizons in microparticle forensics: Actinide imaging and detection of 238Pu and 242mAm in hot particles. *Science Advances* **2021**, *7* (44). DOI: 10.1126/sciadv.abj1175.
(11) Park, J.; Kim, T. H.; Lee, C.-G.; Lee, J.; Lim, S. H.; Han, S. H.; Song, K. Combinatory use of time-of-flight secondary ion mass spectrometry (SIMS) and sector-field SIMS for estimating elemental and isotopic compositions of nuclear forensic samples. *Journal of Radioanalytical and Nuclear Chemistry* **2017**, *311* (2), 1535-1544. DOI: 10.1007/s10967-016-5070-4.
(12) Stephan, T.; Zehnpfenning, J.; Benninghoven, A. Correction of dead time effects in time-of-flight mass spectrometry. *Journal of Vacuum Science & Technology A* **1994**, *12* (2), 405-410. DOI: 10.1116/1.579255.

(13) Lee, J. L. S.; Gilmore, I. S.; Seah, M. P. Linearity of the instrumental intensity scale in TOF-SIMS - a VAMAS interlaboratory study. *Surface and Interface Analysis* **2012**, *44* (1), 1-14. DOI: 10.1002/sia.3761.
(14) Baqué, L. C.; Cabello, F. M.; Viva, F. A.; Corti, H. R. Assessing dead time effects when attempting isotope ratio quantification by time-of-flight secondary ion mass spectrometry. *Biointerphases* **2023**, *18* (6), 061201. DOI: 10.1116/6.0002954.
(15) Kubicek, M.; Holzlechner, G.; Opitz, A. K.; Larisegger, S.; Hutter, H.; Fleig, J. A novel ToF-SIMS operation mode for sub 100 nm lateral resolution: Application and performance. *Appl Surf Sci* **2014**, *289* (100), 407-416. DOI: 10.1016/j.apsusc.2013.10.177 From NLM.
(16) Téllez, H.; Druce, J.; Hong, J. E.; Ishihara, T.; Kilner, J. A. Accurate and Precise Measurement of Oxygen Isotopic Fractions and Diffusion Profiles by Selective Attenuation of Secondary Ions (SASI). *Analytical Chemistry* **2015**, *87* (5), 2907-2915. DOI: 10.1021/ac504409x.
(17) *IONTOF GmbH. SurfaceLab 7.4 Release – Introducing HDR mode for M6 instruments. IONTOF News, 29 February 2024.* https://iontof.com/news-ion-tof-sims-events.html (accessed 2026-09-29).
(18) Debevec, P. E.; Malik, J. Recovering high dynamic range radiance maps from photographs. In Proceedings of the 24th annual conference on Computer graphics and interactive techniques, 1997.
(19) Ievlev, A. V.; Hare, H.; Li, Y.; Kalinin, S. V. PACE-SIMS: Checkpoint-Gated Autonomous SIMS Characterization with AI-Agent Quality Control. *Digital Discovery* **2026**. DOI: 10.1039/D6DD00609D.
(20) Meija, J.; Coplen, T. B.; Berglund, M.; Brand, W. A.; Bièvre, P. D.; Gröning, M.; Holden, N. E.; Irrgeher, J.; Loss, R. D.; Walczyk, T.; Prohaska, T. Isotopic compositions of the elements 2013 (IUPAC Technical Report). *Pure and Applied Chemistry* **2016**, *88* (3), 293-306. DOI: doi:10.1515/pac-2015-0503 (acccessed 2026-09-24).
(21) Baertschi, P. Absolute18O content of standard mean ocean water. *Earth and Planetary Science Letters* **1976**, *31* (3), 341-344. DOI: https://doi.org/10.1016/0012-821X(76)90115-1.
(22) Tyler, B. J. The accuracy and precision of the advanced Poisson dead-time correction and its importance for multivariate analysis of high mass resolution ToF-SIMS data. *Surface and Interface Analysis* **2014**, *46* (9), 581-590. DOI: https://doi.org/10.1002/sia.5543 (acccessed 2026/09/25).

# Supporting Information

## Quantitative Interleaved-Pulse Acquisition Extends the Range of Isotope-Ratio Measurements in ToF-SIMS

Anton V. Ievlev*

*Center for Nanophase Materials Sciences, Oak Ridge National Laboratory, Oak Ridge, Tennessee 37831, United States*

*Email: ievlevav@ornl.gov

### Contents

## S1. Experimental details

### S1.1 Samples and film growth

Thermal $SiO_2$ (100 nm dry oxide on Si) was taken from sealed wafer stock. A $SrTiO_3$(100) single crystal (Alfa Aesar/Thermo Fisher Scientific, product 38507, lot Z15C035; 10 × 10 × 1 $mm^3$, one side polished, Verneuil-grown, undoped) served as the second natural-abundance sample.

The $WO_x$ films were grown at the University of Michigan at room temperature by reactive DC magnetron sputtering (100 W, AJA Orion 8) from a 3-inch W target in Ar/$O_2$ at 40 sccm total flow and 5 mTorr on thermally oxidized Si, with $O_2$ flows of 2, 4, 6 and 8 sccm (5–20% of the total flow). The buried isotopic marker was produced by holding a fixed $^{18}O_2$ flow of 1 sccm during part of the deposition while reducing the $^{16}O_2$ flow by the same amount, so that the total oxygen flow was unchanged; this gives nominal marker fractions of 50, 25, 16.7 and 12.5% of the oxygen supply, assuming isotopically pure $^{18}O_2$. The deposition rate was 5.9–7.1 nm $min^{-1}$; the films are nominally 45 nm thick with the ~15 nm enriched layer in their center, and were capped with ~30 nm $SiN_x$ by plasma-enhanced chemical vapor deposition in a separate tool (Plasmatherm 790).

### S1.2 Acquisition

- Instrument: IONTOF TOF.SIMS 5-NCS, SurfaceLab 7.5, spectrometry (bunched) mode, without the extended-dynamic-range option. $Bi_3^+$ at 30 keV, 30 nA DC, 100 × 100 µm², 128 × 128 pixels, one shot per pixel, 75 µs pulse period (1.22 s per frame measured); pulse widths 1–60 ns.
- Depth profiling in non-interlaced mode: each analysis cycle, containing all phases in sequence (short phase first), is followed by $Cs^+$ sputtering (1 keV, 80 nA, 300 × 300 µm², 3.06 s) and a 2 s pause; one cycle is one layer. Switching between phases costs 122 ± 7 ms per phase slot and scan.
- Charge compensation: low-energy electron flood gun for all measurements; extractor-bias correction in addition for the negative-polarity $SrTiO_3$ measurements.
- Phase sets: $SiO_2$ width sweep 1–60 ns (21 widths, 4 craters) and the four configurations of Table S1; $WO_x$ (4, 6, 9, 9) ns, one frame each; $SrTiO_3$ negative polarity (4, 13.5, 35, 35) ns and positive polarity (9, 13.5, 20, 20) ns, with twin craters of reversed phase order.

### S1.3 Data treatment

- Exported counts carry the vendor single-stop Poisson dead-time correction[S2] applied per pixel; no further correction is applied (a second correction shifts the $SrTiO_3$ ratios by −10 to −25%).
- Peak windows are placed relative to each phase's own apex and matched between the two isotopes (Table S3); $^{17}OH^-$ ($3.81 \times 10^{-4} \times OH^-$, the natural $^{17}O/^{16}O$ ratio) and continuum corrections are applied to the absolute ratios.
- Steady-state layers: $SiO_2$ layers 5–24 (layer 0 surface contamination, layers 1–4 Cs-equilibration ramp with $^{16}O$ rising about 13×); $WO_x$: $k$ from the flank layers below the marker.
- Per-layer precision: on homogeneous $SiO_2$, SD of the residuals about a linear fit over the steady layers, relative to the mean; on $WO_x$, where $f$ varies with depth, SD of the successive differences of ln $f$ divided by √2.

- Natural abundances: IUPAC 2013 representative values; $^{18}O/^{16}O$ referenced to VSMOW, 0.0020052. The source data behind every figure and table, and the exported depth profiles and spectra from which they were computed, are deposited at https://doi.org/10.5281/zenodo.23037549 raw IONTOF project files and analysis scripts are available from the author on request.

## S2. Full performance comparison on thermal $SiO_2$

Table S1 lists the per-crater values behind Table 1. Precision is the detrended per-layer scatter of the $^{18}O/^{16}O$ ratio over layers 5–24 with four analysis frames per cycle; frames and times are scaled to the tuned-pair target of 2.61%, with a timing model of 1.22 s per frame, 122 ms per phase switch, 3.06 s sputter and 2 s pause. Table 1 uses the mean of P1–P3 for the tuned pair (2.6%, 4 frames, 10.2 s) and the 6 ns phase of crater P1 for short-pulse acquisition (9.3%, 51 frames, 67 s). Single widths above 8 ns are biased by $^{16}O$ saturation (deviation from the unsaturated ratio: 8 ns −1%, 9 ns +21 to +25%, 35 ns about +1100%), and the correction of the saturated $^{16}O$ adds 2.5–3.7% per layer of its own noise, so their counting limits (9 ns 3.2%, 35 ns 1.0% at four frames) are lower bounds on the achievable precision. The 9 ns single-width entry of Table 1 is derived from the three 9 ns phases of crater P4 and the 35 ns entry from the single-frame 35 ns phases of P1–P3.

*Table S1.* Per-crater performance on thermal $SiO_2$ (four analysis frames per cycle, layers 5–24). The short-pulse-only row is the 6 ns phase of crater P1 alone, scaled from three to four frames.

| Crater | Configuration (widths, ns; frames) | k | $^{18}O$ long-phase counts per layer | Scatter measured (%) | Scatter counting (%) | Measured / counting | Excess, quadrature (%) | Frames for 2.61% | Analysis (s/layer) | Total (s/layer) |
|---|---|---|---|---|---|---|---|---|---|---|
| P1 | (6, 35) tuned; 3:1 | 39.31 | 2701 | 2.79 | 2.00 | 1.40 | 1.95 | 4.6 | 5.8 | 10.9 |
| P2 | (6, 6, 6, 35) repeated phases; 1:1:1:1 | 42.69 | 2525 | 2.36 | 2.08 | 1.14 | 1.13 | 3.3 | 4.5 | 9.5 |
| P3 | (35, 6) short phase last; 1:3 | 62.76 | 2586 | 2.67 | 2.10 | 1.27 | 1.66 | 4.2 | 5.4 | 10.4 |
| P4 | (6, 9) untuned; 1:3 | 26.53 | 722 | 3.21 | 3.83 | 0.84 | 0.00 | 6.0 | 7.9 | 12.9 |
| P1 short phase | short pulse only, 6 ns (scaled 3 → 4 frames) | — | 92 | 9.32 | 10.51 | 0.89 | — | 51.0 | 62.4 | 67.4 |

## S3. Noise floor of the interleaved reconstruction

*Cross-phase term.* On homogeneous $SiO_2$ the per-layer scatter of a single phase, or of repeated identical phases within a cycle, follows counting statistics within 1.0–1.1× (Table S2). The scatter of a two-width reconstruction is slightly larger than its counting expectation; the difference, added in quadrature, is 1.1–2.0% per layer in the three tuned-pair craters and is quoted as about 1.6% in the main text. The untuned (6, 9) ns crater P4 is counting-dominated (3.8% expected), so the term is not resolved there. Because the term does not scale with counts, it sets the floor on per-layer precision for an interleaved profile.

*Frames per scan versus repeated phases; phase order.* Three frames of one 6 ns phase (P1) and three separate 6 ns phases within the cycle (P2) give the same result: crater-averaged precision 2.70% versus 2.91%, per-layer counting expectation 1.92% versus 1.99%, with the difference in the anchor counts (1375 versus 1184) explained by the lower $^{16}O$ rate of P2. Placing the short phase last (P3) gave about 40% fewer $^{18}O$ counts in the short phase (824 versus 1375) and a poorer crater average (3.48%); the cause was not identified, and the short phase is acquired first throughout.

*Table S2.* Noise-floor data on $SiO_2$ (layers 5–24): per-layer scatter of the reconstruction versus counting expectation and the resulting cross-phase term; spread of identical $^{18}O$ phases within a cycle relative to counting statistics; crater-level anchor counts and precision.

| Crater | Configuration | Measured (%) | Counting (%) | Excess, quadrature (%) | Identical phases, spread / counting | Short-phase $^{18}O$ counts | Crater-averaged precision (%) |
|---|---|---|---|---|---|---|---|
| P1 | (6, 35) tuned, frames 3:1 | 2.79 | 2.00 | 1.95 | — | 1375 | 2.70 |
| P2 | (6, 6, 6, 35) repeated phases | 2.36 | 2.08 | 1.13 | 1.12 (three 6 ns) | 1184 | 2.91 |
| P3 | (35, 6) short phase last, 1:3 | 2.67 | 2.10 | 1.66 | — | 824 | 3.48 |
| P4 | (6, 9) untuned, 1:3 | 3.21 | 3.83 | 0.00 | 1.02 (three 9 ns) | 544 | 4.29 |

*Slow drift of the dose ratio.* Figure S1 shows the per-layer ratio $^{18}O_{long}/^{18}O_{short}$, that is, $k$ layer by layer, and the reconstructed fraction for craters P1 and P4. The reconstructed fraction of the homogeneous oxide drifts by −4.5% (P1) and −4.3% (P4) over layers 5–24; since the film is uniform, this is a drift of the effective dose ratio between the phases of the same size. The per-layer ratio itself is too noisy to show a 4% trend (12–17% per point, set by the 60–80 short-phase $^{18}O$ counts), which is why the drift is best inspected on the reconstructed fraction of a reference region or on the ratio averaged over several layers. Because $k$ enters $f$ as one value per crater, a drift of this size becomes a bias of the same size wherever the layers used for $k$ do not cover the region of interest, which is why the main text recommends inspecting this ratio for trends.

*Alternative tested: normalization to a matrix monitor.* Normalizing each phase per layer to the $Si^-$ matrix signal instead of measuring $k$ from $^{18}O$ was tested on $SiO_2$ in the first measurement session and failed the closure test (the reconstructed ratio did not return the linear short-phase ratio), because $Si^-$ itself is not linear in the long phase and its yield does not track that of $O^-$ through the Cs-equilibration ramp.

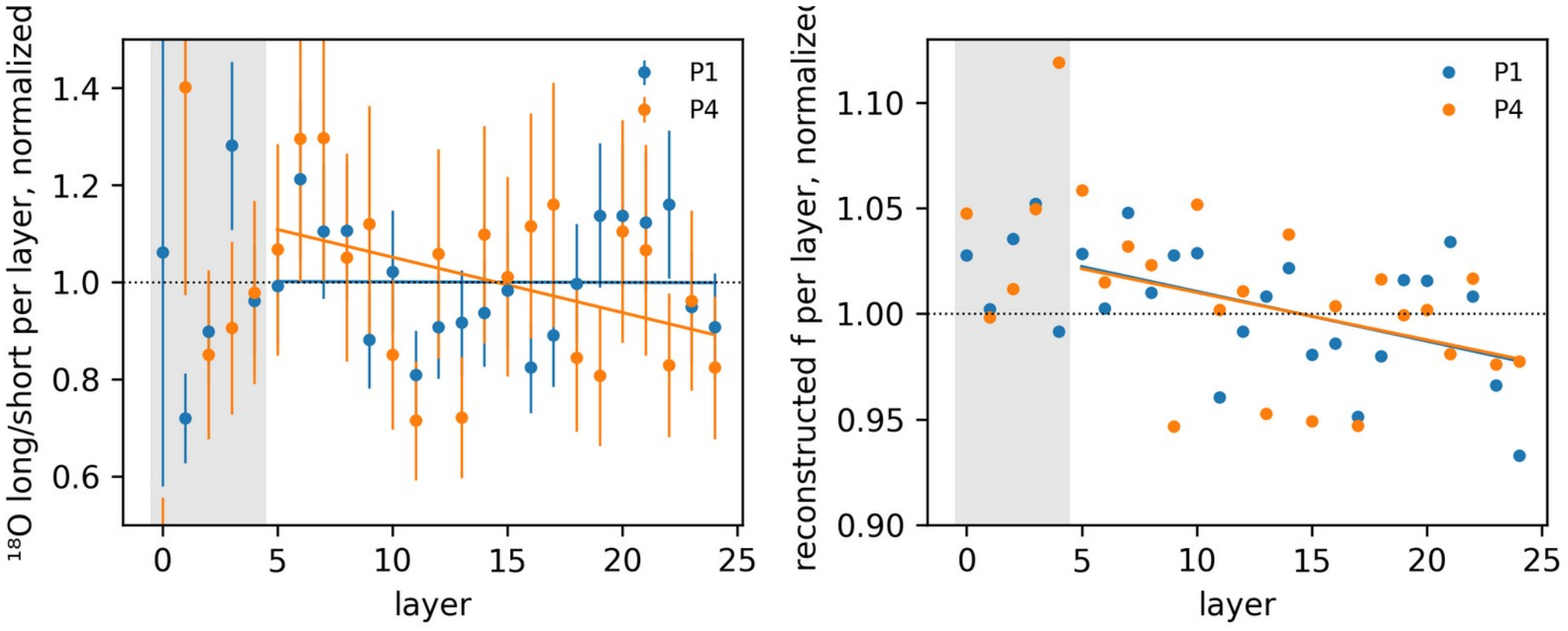

*Figure S1.* Per-layer $^{18}O_{long}/^{18}O_{short}$ ratio (left) and reconstructed $^{18}O$ fraction (right), each normalized to its mean over the steady layers, versus layer for craters P1 (6, 35 ns) and P4 (6, 9 ns) on thermal $SiO_2$. Lines: linear fits over the steady layers. Shaded: surface and Cs-equilibration layers excluded from $k$.

## S4. Peak windows and interference bounds at m/z 18

The $^{18}O^-$ peak is flanked by $^{17}OH^-$ (+7.43 mDa) and $H_2O^-$ (+11.40 mDa). Neither is resolved at the mass resolution of the bunched mode; their contributions are bounded by window counting on per-phase summed spectra. Because the peak apex shifts by about 7 mDa between 1 and 60 ns, all windows are placed relative to the measured $^{18}O^-$ apex of each phase, so the same definitions apply at 6 ns and 35 ns (Table S3).

*Table S3.* Peak windows and measured interference bounds at m/z 18 on thermal $SiO_2$ (per-phase summed spectra of four sweep craters, steady layers 5–24, background = median of 16.30–16.80 u; neighbor windows ±3 mDa around the nominal spacing, counted as upper bounds).

| Quantity | Window | 6 ns (4 craters) | 35 ns (4 craters) |
|---|---|---|---|
| $^{18}O^-$ counts (±4 mDa) | apex ± 4 mDa | 362–454 | 52 400–54 800 |
| $^{17}OH^-/^{18}O^-$ (upper bound) | +7.43 ± 3 mDa | 5.4–10.3% | 1.77–1.84% |
| $H_2O^-/^{18}O^-$ (upper bound) | +11.40 ± 3 mDa | 1.7–2.6% | 0.68–0.74% |
| absolute $^{18}O/^{16}O$ (Table 2) | matched ±4 mDa apex windows on both isotopes; continuum subtracted; $^{17}OH^-$ removed | | |
| per-scan profiles (Table 1, Figure 3) | fixed ±0.1 u export windows (scatter only) | | |

*Alternative tested: doublet decomposition.* Fitting the m/z 18 region as a sum of $^{18}O^-$, $^{17}OH^-$ and $H_2O^-$ peaks of a common line shape, with free amplitudes, was tested in the first measurement session and rejected: the fitted $^{18}O$ amplitude did not reproduce the linear short-phase ratio (closure failure), because at 35 ns the tail of the dominant $^{18}O^-$ peak and the small neighbors are not separable by amplitude alone. The window bounds above are therefore used instead.

## S5. Accuracy: per-crater data

### S5.1 Oxygen at natural abundance

The $^{18}O/^{16}O$ values of Table 2 were derived with one recipe: matched ±4 mDa apex windows on both isotopes, local baselines ($^{16}O$ 15.60–15.85 u; m/z 18 17.70–17.90 and 18.30–18.70 u), $^{17}OH^-$ subtracted, pooled counts over the craters, and the uncertainty taken as the larger of the counting and crater standard errors. Table S4 gives the per-crater values and the window sensitivity; changing the window from ±2.5 to ±6 mDa moves the pooled values by about ±2%, which is the systematic quoted in the main text.

*Table S4.* $^{18}O/^{16}O$ at natural abundance from the 6 ns ($SiO_2$) and 4 ns ($SrTiO_3$, charge compensation on) phases, matched ±4 mDa apex windows, $^{17}OH^-$ subtracted; deviation from VSMOW (0.0020052).

| Set | Crater | $^{18}O/^{16}O$ | Deviation (%) | $^{18}O$ counts |
|---|---|---|---|---|
| $SiO_2$, 6 ns | OFF_r1 | 0.002095 | +4.5 | 788 |
| | OFF_twin | 0.001970 | −1.7 | 663 |
| | OFF_r3 | 0.001978 | −1.3 | 862 |
| | c1 | 0.001950 | −2.8 | 429 |
| | c2 | 0.002096 | +4.6 | 454 |
| | c3 | 0.001877 | −6.4 | 358 |
| | c5 | 0.002210 | +10.2 | 442 |
| | pooled (7 craters) | 0.002023 | +0.9 | counting 1.6%, crater SD 5.6%, SE 2.1% |
| | pooled, windows ±2.5 / ±3 / ±5 / ±6 mDa | | −2.4 / −1.0 / +2.0 / +2.1 | window sensitivity |
| $SrTiO_3$, 4 ns | r1 | 0.002043 | +1.9 | 627 |
| | r2 | 0.002060 | +2.7 | 413 |
| | r3 twin | 0.002024 | +1.0 | 372 |
| | triple | 0.002181 | +8.7 | 429 |
| | pooled (4 craters) | 0.002074 | +3.4 | counting 2.3%, crater SD 3.5%, SE 1.8% |
| | pooled, windows ±2.5 / ±3 / ±5 / ±6 mDa | | +4.0 / +4.2 / +2.9 / +3.8 | window sensitivity |

### S5.2 Sr and Ti isotope ratios

The positive-polarity $SrTiO_3$ data (9, 13.5, 20, 20 ns; three craters) were reduced with the dose chain of S6, one chain per element. $^{88}Sr$ is taken from the 9 ns phase, where it is in the verified linear range, and the weak Sr isotopes from the 20 ns phase; the two phases are linked by $^{86}Sr$, which is linear in both, so that $R(\mathrm{iso}/^{88}\mathrm{Sr}) = (\mathrm{iso}/^{86}\mathrm{Sr})_{20\ \mathrm{ns}} \times (^{86}\mathrm{Sr}/^{88}\mathrm{Sr})_{9\ \mathrm{ns}}$. All Ti isotopes are linear at 20 ns, so their ratios are taken directly from that phase; $^{48}Ti$ measured in the 9 and 20 ns phases serves only as an independent check of the phase response and shows that its pulse-width dependence differs from that of Sr (S6.4). Table S5 gives the per-crater values. The Sr ratios read 2.7–6.3% above natural abundance and the Ti ratios −3.5 to +6.8%; the crater-to-crater reproducibility is 0.3–1.2%. As an internal check (S6.4), $^{88}Sr$ itself is also below 3 corrected counts per pulse at 20 ns and can serve as the bridge; the dose factor it gives differs from the $^{86}Sr$ value by about 3%, and the Sr ratios computed with it lie 2–3% lower ($^{86}Sr/^{88}Sr$ +1.3%, $^{87}Sr/^{88}Sr$ +0.5%, $^{84}Sr/^{88}Sr$ +4.6%). The disagreement is reported, not resolved; it does not affect the Ti ratios. The Ti deviations (+2 to +6%, width-independent, and not saturation: $^{48}Ti$ occupancy about 0.27 at 20 ns) are consistent with minor isobars ($^{46}Ca^+$ on 46; $^{50}Cr^+/^{50}V^+$ on 50) and $^{48}TiH^+$ on 49 (TiH/Ti = 0.62%). $^{87}Rb^+$ is negligible: the $^{85}Rb^+$ monitor is at background.

*Table S5*. $SrTiO_3$ isotope ratios (positive polarity) per crater and mean ± SD, from the $^{86}Sr$ and $^{48}Ti$ chain (Sr: $^{86}Sr$ bridge between 9 and 20 ns, $^{88}Sr$ from 9 ns; Ti: direct 20 ns ratios), as used in Table 2; natural abundances from IUPAC. Last column: values with $^{88}Sr$ as the Sr bridge (internal check, S6.4).

| Ratio | r1 | twin | triple | Mean ± SD | Nat ural | Deviation (%) | $^{88}$Sr bridge (%) |
|---|---|---|---|---|---|---|---|
| $^{84}Sr/^{88}Sr$ | 0.007257 | 0.007192 | 0.007172 | 0.007207 ± 0.000045 | 0.006781 | +6.3 | +4.6 |
| $^{86}Sr/^{88}Sr$ | 0.123994 | 0.122698 | 0.123668 | 0.12345 ± 0.00067 | 0.1194 | +3.4 | +1.3 |
| $^{87}Sr/^{88}Sr$ | 0.087503 | 0.086761 | 0.087026 | 0.08710 ± 0.00038 | 0.08477 | +2.7 | +0.5 |
| $^{46}Ti/^{48}Ti$ | 0.116179 | 0.115818 | 0.116499 | 0.11617 ± 0.00034 | 0.11191 | +3.8 | same |
| $^{47}Ti/^{48}Ti$ | 0.103736 | 0.103256 | 0.103201 | 0.10340 ± 0.00029 | 0.10092 | +2.5 | same |
| $^{49}Ti/^{48}Ti$ | 0.078532 | 0.077398 | 0.079249 | 0.07839 ± 0.00093 | 0.07339 | +6.8 | same |
| $^{50}Ti/^{48}Ti$ | 0.067195 | 0.067737 | 0.068397 | 0.06778 ± 0.00060 | 0.07027 | −3.5 | same |

### S5.3 $WO_x$ markers: linearity and per-layer noise

Each $WO_x$ sample was measured in two craters with the (4, 6, 9, 9) ns set, one frame per phase. The direct short-phase fraction (4 + 6 ns summed) and the (6, 9) ns reconstruction are compared on the marker plateau, where the 9 ns $^{18}O$ reaches 0.4–1.6 counts per pulse (Table S6): the ratio interleaved/direct is 0.993 ± 0.013 over 12.5–50% nominal enrichment with no trend, which is the linearity test of the main text. The same table gives the layer-to-layer noise (successive differences of ln $f$) in the flanks and on the plateau, from which the quantifiable-range estimate of the main text is derived (upper shoulder of S1 r1: 13.3% for the 6 ns phase alone, 3.5% for the reconstruction; plateau 1.9% versus 1.6%). Figure S2 shows the per-layer profiles of one crater per sample. Below the stack, the reconstructed $^{18}O$ fraction of the $SiO_2$ substrate (first 35 substrate layers) is 3.3–9.0% of the marker level, mean 5.8%, flat with depth and uncorrelated with $OH^-$: a natural-abundance reference under an enriched layer is contaminated by recoil mixing and redeposition, so such references are valid only upstream of enriched layers.

*Table S6*. $WO_x$ craters: marker plateau from the direct short-phase fraction (4 + 6 ns) and from the (6, 9) ns reconstruction, 9 ns $^{18}O$ rate on the plateau, and layer-to-layer noise (%) of the 6 ns phase alone / the reconstruction in the upper shoulder (about 0.5–0.8% $^{18}O$), on the plateau, and in the lower shoulder where measured.

| Sample | Crater | *k* | *f* direct (%) | *f* interleaved (%) | Interleaved / direct | 9 ns $^{18}$O (c/p) | Noise upper shoulder | Noise plateau | Noise lower shoulder |
|---|---|---|---|---|---|---|---|---|---|
| S4 (12.5%) | r1 | 21.97 | 11.43 | 11.06 | 0.967 | 0.43 | 22.9 / 5.4 | 3.1 / 2.6 | — |
| | r2 | 19.97 | 11.75 | 11.75 | 1.000 | 0.42 | 23.7 / 5.4 | 5.3 / 2.1 | — |
| S2 (16.7%) | r1 | 20.81 | 16.17 | 16.08 | 0.994 | 0.61 | 15.5 / 2.5 | 3.3 / 1.7 | — |
| | r2 | 20.22 | 16.18 | 16.48 | 1.019 | 0.61 | 22.6 / 3.8 | 4.6 / 2.0 | — |
| S1 (25%) | r1 | 21.30 | 23.51 | 22.75 | 0.968 | 1.10 | 13.3 / 3.5 | 1.9 / 1.6 | 12.4 / 4.4 |
| | r2 | 20.85 | 23.48 | 23.33 | 0.993 | 1.10 | 13.3 / 4.0 | 2.5 / 1.7 | 20.9 / 4.6 |
| S3 (50%) | r1 | 20.25 | 45.25 | 45.19 | 0.999 | 1.60 | 19.5 / 2.9 | 1.8 / 1.3 | 25.0 / 8.3 |
| | r2 | 20.49 | 45.34 | 45.55 | 1.005 | 1.65 | 14.1 / 4.0 | 1.6 / 1.5 | 23.6 / 7.4 |

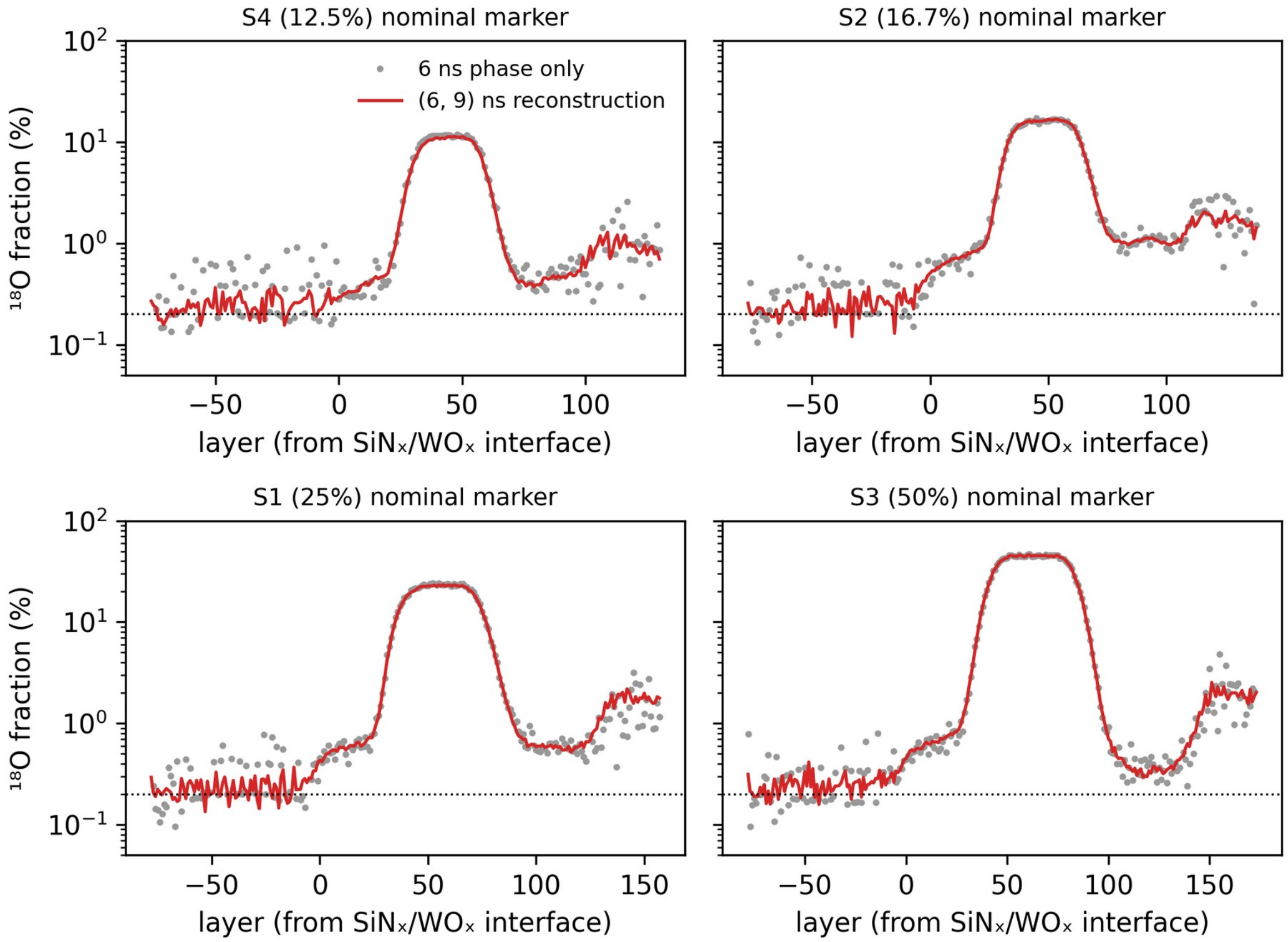

*Figure S2*. Per-layer $^{18}O$ fraction in one crater of each $WO_x$ sample (12.5, 16.7, 25 and 50% nominal marker), from the 6 ns phase alone (grey points) and from the (6, 9) ns reconstruction (red line), on a logarithmic scale; layer index counted from the $SiN_x/WO_x$ interface. Dotted line: natural $^{18}O$ abundance (0.2%). The flank level of about 0.6% is a property of the as-grown films.

## S6. Extension to more than two phases

The two-phase reconstruction of the main text extends to additional pulse widths without changing its logic. Each additional phase is linked to its neighbor by an isotope of the same element that remains linear and spectrally clean in both, and the result is a *dose chain*: each neighboring pair is calibrated exactly as in the two-phase case, and the successive dose factors place all valid measurements on a common scale.

### S6.1 Pairwise dose calibration

Let the phases $j = 1, \ldots, N$ be ordered from the lowest to the highest dose. For two neighboring phases $j$ and $j+1$, choose a bridge isotope $b_j$ that is linear and spectrally clean in both. Its relative phase dose is

$$k_j = \frac{\sum_{l \in A} C_{b,j+1,l}}{\sum_{l \in A} C_{b,j,l}} \tag{S1}$$

where $C_{b,j,\ell}$ is the measured, dead-time-corrected count of the bridge isotope in phase $j$ and layer $\ell$, and the sums run over the calibration layers $A$, typically the steady portion of the crater, in which the relative phase response is stable. Equation S1 is the calibration of main-text Equation 1; no nominal pulse-width ratio or stored attenuation factor enters, and each step of the chain is measured from the sample itself. The cumulative dose of phase $j$ relative to phase 1 is

$$K_1 = 1$$
$$K_j = \prod_{i=1}^{j-1} k_i \quad (j>1) \tag{S2}$$

### S6.2 Reconstruction

For a species $s$, only the phases in which it is both linear and spectrally clean are used; let $J_s$ be this set. Counts from phase $j$ are placed on the phase-1 dose scale through $K_j$. If the species is valid in one phase only, its reconstructed rate is

$$\hat{r}_s = \frac{C_{s,j}}{K_j} \tag{S3}$$

and if it is valid in several phases the valid counts are combined as

$$\hat{r}_s = \frac{\sum_{j \in Js} C_{s,j}}{\sum_{j \in Js} K_j} \tag{S4}$$

that is, total valid counts divided by total relative dose. Equation S4 uses all linear counts of the species and none of its saturated or spectrally compromised ones. Isotope fractions follow from the reconstructed rates in the usual way, for example

$$f = \frac{\hat{r}_{minor}}{\hat{r}_{major} + \hat{r}_{minor}} \quad \text{(S5)}$$

For two phases, Eqs. S1–S5 reduce to main-text Equations 1 and 2: $K_2 = k$, the major isotope is taken from phase 1 and the minor isotope from phase 2 (or from both, Eq. S4).

### S6.3 Design rule

A multiphase acquisition is usable when every neighboring pair of phases is connected by at least one bridge isotope of the same element that remains linear and spectrally clean in both. The practical rule is: (1) use the shortest phase to keep the strongest isotope in the verified linear range; (2) use the longest phase that preserves the required spectral separation for the weakest isotope; (3) insert intermediate phases only when no single long phase can remain valid across the full isotope-ratio range; (4) ensure that each neighboring pair shares at least one valid bridge isotope. The dose steps should be as large as this overlap allows, because larger steps reduce the number of phases needed.

### S6.4 Internal checks

When more than one isotope is linear in the same pair of phases, each provides an independent estimate of the pairwise dose factor, and their agreement is an internal consistency check. Disagreement indicates that an assumption should be re-examined: partial detector saturation, an unresolved spectral interference, a pulse-width-dependent response that differs between mass ranges, or drift of the relative phase response through the crater. Dose calibration is therefore performed separately for different elements when required. In the $SrTiO_3$ data the 20 ns dose relative to 9 ns is 4.64 for the Sr isotopes and 5.14 for the Ti isotopes, a difference of about 10%, so separate Sr and Ti chains are used; within the Sr chain, $^{86}$Sr and $^{88}$Sr as alternative bridges between 9 and 20 ns give dose factors that differ by about 3% (S5.2). The same principle applies to drift within a crater: the bridge-isotope ratio should be checked over the region of interest, preferably as a running or region-averaged value when individual layers are too noisy (Figure S1).

### S6.5 Examples

*$WO_x$, three pulse widths.* The $WO_x$ craters were measured at 4, 6 and 9 ns. $^{16}$O is linear at 4 and 6 ns but saturates at 9 ns; $^{18}$O is linear at all three widths and links both pairs. The chain measured from $^{18}$O on the 25% nominal marker crater is $k_{4\to6} = 4.31$ and $k_{6\to9} = 10.66$, so the three phases are on one scale; $^{16}$O is taken from the 4 and 6 ns phases combined (Eq. S4), $^{18}$O from all three. On the plateau the three-phase reconstruction agrees with the (6, 9) ns reconstruction to 0.999, and it lowers the marker noise from 1.6 to 1.3% where the short-phase $^{16}$O counts limit the result, leaving the dilute flanks unchanged (Table S7).

*$SrTiO_3$, element-specific chains.* The positive-polarity $SrTiO_3$ measurements (9, 13.5, 20, 20 ns) cover the abundant and the weak Sr and Ti isotopes. For Sr, $^{86}$Sr links the 9 and 20 ns phases, with $^{88}$Sr taken from 9 ns and the weak Sr isotopes from 20 ns. For Ti, all isotopes remain linear at 20 ns, so their ratios are obtained directly from that phase; $^{48}$Ti measured across phases provides an independent check of the phase response and shows that its pulse-width dependence differs from that of Sr. This is why a bridge isotope should belong to the same element as the isotopes being reconstructed, and why no dose chain is built when a direct linear measurement is available. Not every acquired phase has to enter the final chain: the 13.5 ns phase was omitted because the $^{86}$Sr bridge spans the larger 9 → 20 ns step directly, and the two 20 ns phases were combined; redundant phases can be omitted or retained as internal consistency checks. As a negative control, applying the dead-time correction a second time to the already corrected exports shifts all Sr and Ti ratios by −10 to −25%.

The multiphase extension is thus not a different method from the two-phase case but a repetition of the same operation: measure, link neighboring phases, rescale, use only valid signals. Each pair of phases is calibrated from the data themselves, each isotope contributes only where its response is linear and spectrally clean, and additional phases extend the calibrated dose chain to a wider intensity range.

*Table S7.* Three-phase reconstruction of the $WO_x$ 25% marker crater (4, 6, 9 ns): validity of each isotope per phase (maximum corrected rate, counts per pulse), the dose chain measured from $^{18}$O, and the layer-to-layer noise by region for the three-phase reconstruction, the (6, 9) ns pair and the 6 ns phase alone.

| Quantity | Value |
|---|---|
| $^{16}$O validity at 4 / 6 / 9 ns | linear (≤ 0.17) / linear (≤ 0.71) / saturated (up to 5.7) |
| $^{18}$O validity at 4 / 6 / 9 ns | linear (≤ 0.03) / linear (≤ 0.11) / linear (≤ 1.1) |
| dose chain from $^{18}$O: $k_{4\rightarrow 6}$, $k_{6\rightarrow 9}$ | 4.31, 10.66 |
| plateau fraction, three-phase / (6, 9) pair | 0.999 |
| noise, three-phase / (6, 9) pair / 6 ns alone: upper shoulder | 3.8 / 3.5 / 13.3% |
| noise: marker plateau (22.5%) | 1.3 / 1.6 / 1.9% |
| noise: lower shoulder | 4.3 / 4.4 / 12.4% |

## S7. Comparison with the vendor high-dynamic-range (HDR) mode

SurfaceLab (IONTOF GmbH) offers, from version 7.4, a high-dynamic-range (HDR) mode that acquires two datasets at different pulse durations (or currents) of the same field of view in one automated run and merges them, by analogy with HDR photography,[S4] into images and profiles without saturated pixels at maintained sensitivity.[S1] We are not aware of a peer-reviewed characterization of the mode. Because it acquires the same kind of two-width data as the method of the main text, the two can be compared on identical counts: the component datasets exported from an HDR run were reconstructed with the dose factor measured in the crater and validity selection, and set against the vendor merge of the same run.

### S7.1 Thermal $SiO_2$, tuned (6, 35) ns widths

Three positions on the thermal $SiO_2$ of the main text were measured in the HDR mode with the 6 and 35 ns widths of Table 1 (one frame per phase, 30 layers; otherwise the conditions of S1.2), and the 6 ns, 35 ns and HDR-merged profiles were exported from the same run. $^{16}O^-$ has a raw occupancy of 0.47–0.50 at 6 ns, linear after correction as in Figure 1c, and 0.997 at 35 ns, i.e. it is fully saturated; $^{18}O^-$ is linear at both widths (35 ns dead-time factor 1.07). The dose factor from $^{18}O^-$ is $k$ = 101.6 ± 4.0 per frame, somewhat lower than the per-frame equivalent of the Table S1 measurements (39.3 per scan, about 118 per frame), which itself illustrates why the dose factor is measured in each crater rather than stored; the nominal duration ratio is 5.8.

*Merge rule*. For $^{18}O^-$, which is linear in both datasets, the HDR output equals the sum of the corrected 6 and 35 ns counts within 0.1%; for the saturated $^{16}O^-$ it is 0.37–0.40 of the dose-consistent value, 6 ns $^{16}O \times (1 + k)$. Expressed as a rule, HDR $^{18}O$ = sum of the phases and HDR $^{16}O = c \times$ 6 ns $^{16}O$ with $c$ = 39.1–39.7, a multiplier that does not follow the measured dose factor. The datasets are thus superposed with a global weight for the saturated channel, not converted to a common dose scale.

*Result*. On the exported profile windows (identical for all three quantities), the vendor merge reads 2.6 times natural abundance (+173%) at every position, whereas the reconstruction of the same data returns +4.9 ± 5.5% and, by construction, the crater average of the linear 6 ns phase (Table S8, Figure S3). The per-layer scatter (layers 5–29) is 17–21% for the 6 ns phase alone and 2.4–3.3% for both the reconstruction and the HDR profile: the merge has the precision of the long phase and a 2.6-fold bias; the reconstruction has the same precision and no bias.

*Table S8.* Thermal $SiO_2$ in the vendor HDR mode with the tuned (6, 35) ns widths, three positions: $^{18}O/^{16}O$ deviation from natural abundance (VSMOW) for the 6 ns phase alone, the dose-calibrated reconstruction of the two phases and the vendor HDR merge, all on identical exported profile windows. Per-layer scatter over layers 5–29.

| Position | 6 ns alone | Reconstruction | Vendor HDR | HDR / reconstruction |
|---|---|---|---|---|
| 1 | +8.5% | +8.5% | +178% | 2.57× |
| 2 | +7.7% | +7.7% | +174% | 2.54× |
| 3 | −1.4% | −1.4% | +166% | 2.69× |
| mean | +4.9% | +4.9 ± 5.5% | +173% | 2.60× |
| per-layer scatter | 17–21% | 2.4–3.3% | 2.4–3.3% | |

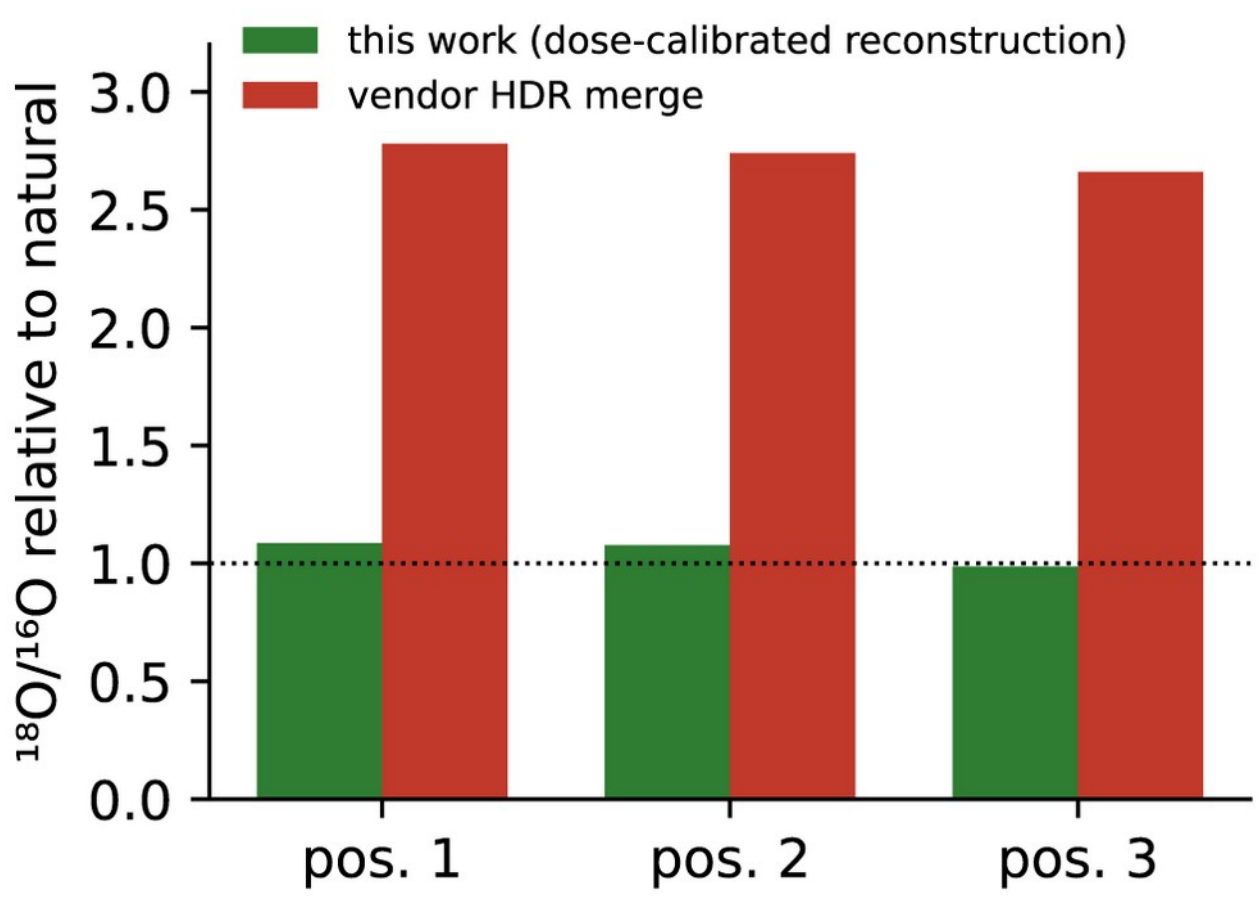

*Figure S3.* $^{18}O/^{16}O$ relative to natural abundance at the three $SiO_2$ positions: the dose-calibrated reconstruction (green) and the vendor HDR merge (red), both from the same exported profiles. Dotted line: natural abundance.

### S7.2 $^{18}O$-labelled $HfO_2$ film, imaging mode

An earlier test on an $^{18}O$-labelled $HfO_2$ film ($SiN_x$ cap, Si substrate; imaging mode, $Bi_3^+$, 10 and 100 ns, 128 × 128 pixels, $Cs^+$ 1 keV, 120 layers) gave the same picture with a different dose ratio ($k$ = 13.4 against a nominal 10). Unsaturated channels ($SiN^-$, $^{18}O^-$, $Si^-$, $HfO^-$) are summed within 0.1–1%; the saturated $^{16}O^-$ (100 ns dead-time factor 1.33–1.37) is retained at 0.40 of the dose-consistent value with the automatic coefficients (0.5551/0.4441), 0.78 with the coefficients set to $k$, and 0.27 reversed, so the coefficients act only where a channel is saturated and no setting reaches the dose-consistent value. The merge overestimates $^{18}O/^{16}O$ in the labelled oxide by 2.5× (automatic), 1.4× (set to $k$) and 4× (reversed), whereas the reconstruction of the same datasets agrees with the linear 10 ns ratio within 1–5% at far lower noise. In the nitride cap and the substrate the 100 ns peaks are too broad for the matrix (the $^{16}O^-$ window most likely includes $NH_2^-$), so both long-pulse results deviate there; no combination scheme corrects a long phase that violates the peak-width rule of the main text, which is why S7.1 uses the tuned widths.

### S7.3 Interpretation

A photographic HDR merge converts each exposure to a common scale, weights each by its reliability with zero weight for clipped pixels, and averages; with Poisson statistics and weights proportional to exposure, the merge of unsaturated data reduces to the sum of the raw counts,[S5] which is what the vendor output shows. This merging strategy does not preserve quantitative ion ratios when one component dataset is saturated, for three reasons: the relative exposure is not measured in each analysis but taken from global coefficients, while the measured dose ratio differs from the nominal duration ratio (about 100 versus 5.8 here), differs between species of different mass (S6.4) and drifts within a measurement (S3); saturated data remain in the sum with a weight instead of being replaced by the linear short-pulse signal; and tuning the coefficients until a known ratio is reproduced requires knowing the answer. The mode fulfils its stated purpose, saturation-free images and profiles with good statistics, and its precision on $SiO_2$ is that of the long phase. What makes the same data quantitative is the two steps of the main text: a dose factor measured in every crater from a species linear in both phases, and validity selection.

## References

Figures, tables and equations without the prefix S refer to the main text.

(S1) IONTOF GmbH. SurfaceLab 7.4 released: new HDR mode. News announcement, 29 February 2024. https://www.iontof.com (accessed September 2026).

(S2) Stephan, T.; Zehnpfenning, J.; Benninghoven, A. Correction of dead time effects in time-of-flight mass spectrometry. *J. Vac. Sci. Technol. A* **1994**, *12*, 405–410.

(S3) Ievlev, A. V.; Hare, H.; Li, Y.; Kalinin, S. V. PACE-SIMS: Checkpoint-gated autonomous SIMS characterization with AI-agent quality control. *Digital Discovery* **2026**. *DOI: 10.1039/D6DD00609D.*

(S4) Debevec, P. E.; Malik, J. Recovering high dynamic range radiance maps from photographs. In *Proceedings of SIGGRAPH 97*; ACM: New York, 1997; pp 369–378. DOI: 10.1145/258734.258884.

(S5) Granados, M.; Ajdin, B.; Wand, M.; Theobalt, C.; Seidel, H.-P.; Lensch, H. P. A. Optimal HDR reconstruction with linear digital cameras. In *2010 IEEE Conference on Computer Vision and Pattern Recognition*; IEEE: Piscataway, NJ, 2010; pp 215–222. DOI: 10.1109/CVPR.2010.5540208.